\documentclass[10pt,conference]{IEEEtran}
\IEEEoverridecommandlockouts
\usepackage{tikz}
\usetikzlibrary{positioning,fit,arrows.meta,shapes.geometric}
\usepackage{cite}
\usepackage{amsmath,amssymb,amsfonts}
\usepackage{algorithmic}
\usepackage{graphicx}
\usepackage{textcomp}
\usepackage{xcolor}
\usepackage[most]{tcolorbox}
\def\BibTeX{{\rm B\kern-.05em{\sc i\kern-.025em b}\kern-.08em
    T\kern-.1667em\lower.7ex\hbox{E}\kern-.125emX}}
\usepackage{xspace}
\usepackage{graphicx}
\usepackage{subfig}
\usepackage{booktabs}
\usepackage{subcaption}
\usepackage{url}
\usepackage[hidelinks]{hyperref}
\newcommand{\NeuroSTAR}{\textsc{NeuroSTAR}\xspace}
\newcommand{\LTL}{\ensuremath{\mathrm{LTL}}\xspace}
\newcommand{\LTLf}{\ensuremath{\mathrm{LTL}_{f}}\xspace}
\newcommand{\LTLorf}{\ensuremath{\mathrm{LTL}_{(f)}}\xspace}

\newcommand{\ltlf}{\LTLf}
\newcommand{\buchi}{B\"{u}chi\xspace}
\newcommand{\AdaRule}{\textsc{AdaRule}\xspace}

\newcommand{\ConfoNL}{\textsc{ConfoNL}\xspace}
\newcommand{\ConfoNLFiltered}{\textsc{ConfoNL}$_{f}$\xspace}
\newcommand{\ConfoNLStrict}{\textsc{ConfoNL}$_{st}$\xspace}

\newcommand{\SynthNL}{\textsc{SynthNL}\xspace}
\newcommand{\SynthNLFiltered}{\textsc{SynthNL}$_{f}$\xspace}
\newcommand{\SynthNLStrict}{\textsc{SynthNL}$_{st}$\xspace}

\newcommand{\LangNL}{\textsc{LangNL}\xspace}
\newcommand{\LangNLFiltered}{\textsc{LangNL}$_{f}$\xspace}

\newcommand{\SpecNL}{\textsc{SpecNL}\xspace}
\newcommand{\SpecNLFiltered}{\textsc{SpecNL}$_{f}$\xspace}

\newcommand{\VALaw}{\textsc{VALaw}\xspace}
\newcommand{\valaw}{\VALaw}

\usepackage{tikz}
\usetikzlibrary{positioning,calc,fit,backgrounds}
\newcommand{\tLTLf}{\mathrm{LTL}_{f}}

\newcommand{\ap}[1]{\ensuremath{\texttt{#1}}}
\newcommand{\temporal}[1]{\ensuremath{\mathcal{#1}}}

\newcommand{\ignore}[1]{}
    
\begin{document}

\title{\NeuroSTAR: Automata-guided Neuro-symbolic Specification Formalization}

\author{
\IEEEauthorblockN{Joy Saha}
\IEEEauthorblockA{
\textit{University of Virginia} \\
Charlottesville, VA, USA \\
daa7mv@virginia.edu
}
\and
\IEEEauthorblockN{Trey Woodlief}
\IEEEauthorblockA{
\textit{William \& Mary} \\
Williamsburg, VA, USA \\
woodlief@wm.edu
}
\and
\IEEEauthorblockN{Sebastian Elbaum}
\IEEEauthorblockA{
\textit{University of Virginia} \\
Charlottesville, VA, USA \\
selbaum@virginia.edu
}
\and
\IEEEauthorblockN{Matthew B. Dwyer}
\IEEEauthorblockA{
\textit{University of Virginia} \\
Charlottesville, VA, USA \\
matthewbdwyer@virginia.edu
}
}

\maketitle

\begin{abstract}
Automated translation of natural language (NL) descriptions into Linear Temporal Logic over finite traces (\LTLf) is a prerequisite for automated formal verification of a system's dynamic behavior. Several LLM-based methods have recently shown potential for this task. 
However, they struggle with the nuance of natural language descriptions, which can lead LLMs to only partially capture the intended meaning.
To address this limitation, we propose
\NeuroSTAR (Automata-guided Neuro-symbolic Specification Formalization), an  NL-to-\LTLf framework that builds on two insights.
First, it leverages multiple generators to obtain diverse \LTLf candidates. 
Second, it uses an automata-theoretic semantic comparison based on DFA traces to identify behavioral disagreements
that
guide formula refinement.
We evaluate \NeuroSTAR and show that it  improves NL-to-\LTLf translation 
performance by 8-18 percentage points relative to the prior state-of-the-art (SoTA) on unambiguous benchmarks. We further study its applicability to a body of driving law text, a complex, realistic, and reference-free domain critical for autonomous-vehicle specification. This study shows that \NeuroSTAR can capture the necessary temporal semantics in 83.9\% of the driving law sections, which
demonstrates the effectiveness of automata-guided reference-free refinement in formalization. 
\end{abstract}

\begin{IEEEkeywords}
Temporal logic, automated formalization, difference automata
\end{IEEEkeywords}

\section{Introduction}
\label{sec:introduction}
Temporal logics have been studied for nearly 5 decades~\cite{pnueli1977temporal}.
Research into leveraging temporal logic specifications
for system verification~\cite{baier2008principles},
monitoring~\cite{bauer2011runtime},
and synthesis~\cite{he2019efficient} has moved from theory to 
application in domains like:
planning~\cite{kress2009temporal},
robotics~\cite{menghi2019specification,vazquez2024robotics},
automotive~\cite{fainekos2012verification,toledo2024specifying}, and
aerospace~\cite{luppen2022elucidation}.

Formal specifications in a suitable logic for such tasks are not always available; instead, 
behavior is typically described only in natural language~\cite{bogli2025temporal,rozier2016specification}.
The software engineering community has put forth many efforts to support
the development of formal temporal specifications, e.g.:
documenting recurring specification patterns~\cite{dwyer1999patterns},
mapping to them from structured natural language~\cite{giannakopoulou2020generation},
mining them from system traces~\cite{lemieux2015general}, and
learning them from examples with side-constraints~\cite{zhang2025constrained}.

The advent of large language models (LLMs) has led to a surge of interest in exploring how
they might facilitate the formalization of temporal requirements~\cite{ma2025specgen,9604679,10684640,hu2026adarule,sundarsingh2026conformal,nl2spec,rabiei2025ltlcodegencodegenerationsyntactically,mavrogiannis2024cook2ltltranslatingcookingrecipes,10161125,ma2025bridging}. 
While these techniques have shown promise, none exploit the fact that
the target temporal logic has a well-defined semantics that is amenable to symbolic encoding and algorithmic manipulation.
Recent work~\cite{danso2026syntax} has shown that while LLMs can often produce syntactically
correct formalizations, they are much less effective in capturing semantic constraints.

In this paper, we present \NeuroSTAR, a
neuro-symbolic approach that  translates temporal requirements expressed in
natural language (NL) into linear temporal logic over finite traces (\LTLf).
Like other approaches, \NeuroSTAR leverages LLMs to translate natural language requirements to temporal logic specifications
over a given set of \textit{atomic propositions} (APs) that capture observations about
the problem domain.
Unlike prior work, \NeuroSTAR converts temporal logic specifications 
to automata and uses
them to generate symbolic representations of sets of traces on which
alternative specifications disagree.   
We use those disagreements to drive an iterative refinement process
where LLMs are tasked with minimizing those disagreements to \textit{converge} on a final specification.

To evaluate \NeuroSTAR, we began by replicating the prior SoTA, \AdaRule~\cite{hu2026adarule}. 
During the replication process, we
noticed inconsistencies and anomalies
in the datasets used to train and evaluate both \AdaRule and the baselines it
compared with.   
In \S~\ref{sec:bencheval}, we present a detailed analysis of ambiguity and other inconsistencies
of those datasets. 
We also evaluate the extent to which these datasets capture the interpretive challenges present in real-world temporal requirements. 

Our evaluation of \NeuroSTAR demonstrates the value of the neuro-symbolic design, and we find that it establishes a new 
state of the art on benchmarks used in prior work~\cite {hu2026adarule,sundarsingh2026conformal,crick2023nl2ltlsynthetic,liu2023groundingcomplexnaturallanguage,nl2spec}.
Across six variants of four datasets from prior work~\cite {hu2026adarule}, \NeuroSTAR achieves the best or joint-best performance on five of them.
We further explore the applicability of \NeuroSTAR on a corpus
of natural language requirements that codify the Virginia driving code~\cite{virginia_code_title46_2_2025}, \valaw, which are challenging, semantically complex,
layered natural language statements, with cross-references.
On this corpus, \NeuroSTAR produces formalizations that satisfy 
a necessary condition for correctness for $89$ of $106$ reviewed laws, correctly modeling obligations, exceptions, and interactions among multiple road users.

In summary, the primary contributions of this paper are:
(1) a comprehensive audit of widely used NL-to-\LTLorf\footnote{ \LTLorf refers to \LTL \textit{and/or} \LTLf when interchangeable; see \S~\ref{sec:temporal_logics}.} benchmarks, exposing their limitations;
(2) identification of the gap between existing NL-to-\LTLorf benchmarks and real-world specifications;
(3) presentation of \NeuroSTAR, a training-free neuro-symbolic framework that combines neural generation with symbolic automata-guided reasoning for NL-to-\LTLf formalization;
(4) an evaluation demonstrating that \NeuroSTAR improves on the prior state-of-the-art; and
(5) a reference-free study on the Virginia driving code demonstrating \NeuroSTAR's ability
to handle complex real-world specifications.

\section{Background and Related Work}




\subsection{Temporal Logics}
\label{sec:temporal_logics}
We focus on Linear Temporal Logic (\LTL) and the related Linear Temporal Logic over Finite Traces (\LTLf)~\cite{de2013linear}.
\NeuroSTAR targets \LTLf 
due to its use for cyber-physical systems discussed below, while prior work targeted \LTL
~\cite{hu2026adarule}.
We explore the impact of this difference
in \S~\ref{sec:ambiguous_nl}.

\LTL and \LTLf have the same syntax but differing semantics.
A formula under \LTLorf is interpreted over a sequence of evaluations of \textit{atomic propositions} (APs), collectively referred to as a \textit{trace},
and produces a boolean result as to whether the trace is in the \textit{language} of the formula.
In formalizing natural language specifications, each AP refers to a boolean-valued event, action, or measurement with defined or inferred semantics; e.g., the \ap{is\_off\_road} AP is true iff the autonomous vehicle is off the drivable roadway.
A formula connects the APs with the usual boolean operators $(\wedge,\vee,\neg,\implies)$ along with the temporal operators \textit{Next} $(\temporal{X}~\phi)$, \textit{Until} $(\phi_1\mathbin{\temporal{U}}\phi_2)$, \textit{Always} $(\temporal{G}~\phi)$, and \textit{Eventually} $(\temporal{F}~\phi)$.
A formula in \LTL is evaluated over an \textit{infinite}-length trace,
while \LTLf is evaluated over a \textit{finite}-length trace.
Formulae in \LTLorf can be compiled into \textit{automata} that enable efficient and automated analysis and evaluation of traces.
Due to their operation over infinite-length traces, \LTL formulae compile to \buchi automata~\cite{vardi1994reasoning}, while \ltlf formulae compile to Deterministic Finite Automata (DFAs)~\cite{de2013linear}.
Researchers have developed a rich body of software that supports the automatic compilation of \LTLorf automata~\cite{duret.22.cav,fuggitti-ltlf2dfa}.
Many \LTL and \LTLf formulae have intuitively similar behavior.
For example, the formula $\temporal{G}(\phi)$ in \LTL is read as ``$\phi$ always holds,'' while in \LTLf it is read as ``$\phi$ always holds \textit{during the trace}.''
However, the differing semantics lead to differing expressibility;
e.g., the expression $\phi \wedge \temporal{G}(\phi \implies (\temporal{X}~\neg\phi) \wedge (\temporal{X}\temporal{X}~\phi))$ is read as ``$\phi$ holds every other time step'' in \LTL but
is the empty language
in \LTLf.

We target \LTLf 
due to its demonstrated practical utility in formal specifications for cyber-physical systems 
due to the nature of finite-length missions
~\cite{zhang2025eteltlf,WOODLIEF2025103252,woodlief2025scene,2024iv...conf..219L,9561297,gitelson2025maximaladaptationminimalguidance}.
Prior work has utilized the Virginia driving laws~\cite{virginia_code_title46_2_2025} as 
specifications in studying runtime monitoring for autonomous vehicles~\cite{WOODLIEF2025103252,woodlief2025scene};
this serves
as the basis for analysis in \S~\ref{sec:va-law}.



\subsection{Natural Language to Temporal Logic Translation}
Natural language (NL) to temporal logic translation aims to convert an informal requirement into a specification in a formal logic such as \LTL or \LTLf. Driven by the practical need for formalization, work in this area has progressed through three broad generations.  
Early work~\cite{5362112} treats NL-to-\LTL as a classification problem over a fixed library of formalization patterns and uses machine learning classifiers to identify Dwyer property patterns~\cite{dwyer1999patterns} in NL, leaving formula synthesis as future work. 
A second generation of approaches~\cite{hahn2022formalspecificationsnaturallanguage,chen2024nl2tltransformingnaturallanguages,9812169} fine-tune encoder–decoder transformer language models, such as T5~\cite{raffel2023exploringlimitstransferlearning}, on paired NL-to-\LTL datasets.
This approach is further strengthened by incorporating grammar-forced decoding
during
generation~\cite{english2025grammarforcedtranslationnaturallanguage}.
However, due to the scalability limitations of these approaches and the rise of large language models (LLMs) in software engineering tasks, recent work has shifted toward LLM-based methods. 
LLM-based approaches either fine-tune a pretrained LLM~\cite{10161125}, learn patterns from data~\cite{hu2026adarule,10684640}, or use reference-free LLM-based strategies~\cite{nl2spec,mavrogiannis2024cook2ltltranslatingcookingrecipes,rabiei2025ltlcodegencodegenerationsyntactically,sundarsingh2026conformal,ma2025bridging,fuggitti2023nl2ltl} for this task. 


Despite this progress, existing LLM-based strategies suffer from distinct vulnerabilities. Data-dependent methods (fine-tuning and pattern learning) are bounded by the quantity and quality of paired NL-to-\LTL datasets. 
Such annotations can be noisy or inconsistent, and in specialized domains such as legal text, a sufficiently large labeled dataset often does not exist at all.
Reference-free LLM-based approaches, although free from this data dependency, instantiate a \emph{single-generator} paradigm---one model, one decoder, one perspective---leaving no external check against the biases and blind spots inherited from its training. 
Furthermore, no approach leverages the rich automata-theoretic machinery underlying temporal logic to reason about the semantics of the generated formula.
Thus, motivated by these observations, we design \NeuroSTAR as a reference-free, multi-generator-based NL-to-\LTLf framework that integrates automata-theoretic reasoning at its core.

\section{Benchmark Study}
\label{sec:bencheval}
As noted in \S~\ref{sec:introduction}, while replicating \AdaRule~\cite{hu2026adarule}, the SoTA baseline we compare against in \S~\ref{sec:expII}, we identified weaknesses in the chosen evaluation datasets used in prior work: ambiguity, test-train contamination, trivial true/false formulae, incorrect ground truth, and simplicity.
Ambiguity in the NL specifications can preclude formalization as multiple plausible interpretations exist; in such cases, labeling one as ``correct'' adds information not present in the NL.
Simplicity in the corpus of NL-\LTL pairs covered also limits generalization as these are not representative of real-world specifications. 
These weaknesses would directly impact the construct validity and external validity of our later evaluations.
Below, we design two research questions to quantify each of these weaknesses through manual analysis of the four datasets studied:
\ConfoNL~\cite{sundarsingh2026conformal},
\SynthNL~\cite{crick2023nl2ltlsynthetic},
\LangNL~\cite{liu2023groundingcomplexnaturallanguage}, and
\SpecNL~\cite{nl2spec}.
In the process, we will develop refined versions of the datasets that we will use in our later evaluation in \S~\ref{sec:expII}, helping ameliorate these threats.

\begin{itemize}
    \item \textbf{RQ1:} To what extent are existing NL-to-\LTL benchmarks appropriate for evaluation of \NeuroSTAR? Do they contain ambiguous natural-language specifications or incorrect ground-truth \LTL formulae? What is the impact of switching from \LTL to \LTLf? 
    \item \textbf{RQ2:} To what extent do existing
    benchmarks capture the interpretation challenges 
    of
    real-world specifications?
\end{itemize}

\subsection{RQ1: Benchmark Ambiguity}
\label{sec:ambiguous_nl}

First, as \NeuroSTAR targets \LTLf, to ensure fair comparison in the evaluation, we remove any NL-\LTL pairs that do not have equivalent semantics under \LTLf.
We also remove pairs where the \LTLf is incorrect or is trivially true/false.
Together, these remove
18 (5.7\%) NL-\LTL pairs across the original four datasets.
\AdaRule uses a train-test split of the data
for developing rules and during evaluation, respectively.
However, 
we found that for the \LangNL dataset, 76 (76\%) of the test NL-\LTL pairs were exactly duplicated in the train and test splits.
We remove these from the evaluation to prevent train-test contamination.
We denote these filtered datasets as \ConfoNLFiltered, \SynthNLFiltered, \LangNLFiltered, and \SpecNLFiltered.

Next, we identify two sources of ambiguity in the datasets: \textit{temporal ambiguity} and \textit{anaphoric ambiguity}.
These sources of ambiguity only impact \ConfoNLFiltered and \SynthNLFiltered; we will address these in the creation of \ConfoNLStrict and \SynthNLStrict as described next.
\textit{Temporal ambiguity} arises from the multiple interpretations of timing in NL.
One consideration is how \textit{strict} the temporal language should be interpreted.
Consider the following NL specification from \ConfoNLFiltered:
``First go to \ap{AP1}, then proceed to \ap{AP2} and stay there.''
In a strict interpretation, the word ``first'' means that \ap{AP1} must strictly precede \ap{AP2}; i.e., \ap{AP2} cannot hold before the first \ap{AP1}.
However, under a non-strict interpretation, the word ``first'' could be removed without altering the meaning; i.e., as long as \ap{AP1} happens and then in the future \ap{AP2} holds indefinitely, the specification holds, regardless of if \ap{AP2} held at some point before the first occurrence of \ap{AP1}.
These two interpretations lead to differing \LTLorf.
\begin{align*}
\text{Strict:}&~~\neg\,\ap{AP2}\mathbin{\temporal{U}}(\ap{AP1}\mathbin{\wedge} \temporal{F}\temporal{G}~\ap{AP2})\\
\text{Non-strict:}&~~ \temporal{F} (\ap{AP1}\mathbin{\wedge}\temporal{F} (\ap{AP2} \wedge \temporal{G}~\ap{AP2}))
\end{align*}

From inspecting the labels, we find that \ConfoNLFiltered and \SynthNLFiltered consistently favor the non-strict interpretation of the NL---the non-strict variant above is the correct label as defined in \ConfoNLFiltered.
However, we argue that, in the context of specification formalization, the maxim of quantity~\cite{grice1975logic} favors the strict interpretation---we must assume the author chose the word ``first'' deliberately to alter the semantics of the specification, i.e., a strict interpretation.
Given that the dataset consistently favors the non-strict interpretation, \AdaRule benefits substantially from its training process in which it can identify this pattern that is not clear from the NL.
To remove this inherent advantage and improve generality to the reference-free case, for the two datasets containing such ambiguity, \ConfoNLFiltered and \SynthNLFiltered, we generate strict versions of the datasets by removing\footnote{There may be multiple strict versus non-strict ambiguities within a specification, making the choice between alternatives less clear. To avoid adding our own biases, we remove rather than edit temporal ambiguities.} the pairs where the strict 
and non-strict interpretations diverge.
We denote these datasets as \ConfoNLStrict and \SynthNLStrict; between the two datasets, this removed 60 pairs (31.08\%).

Natural language specifications suffer from \textit{anaphoric ambiguity} when a word that references another part of the sentence is ambiguous in its target~\cite{jafari2023anaphoric}. 
Consider the following NL specification from the \ConfoNLFiltered dataset:
``Do not visit \ap{AP1} before reaching \ap{AP2}, and after that, proceed to enter and remain there.''
In this sentence, it is ambiguous whether ``there'' refers to \ap{AP1} or \ap{AP2}; however, inspecting the \LTL label $\neg\ap{AP1}~\temporal{U}~\ap{AP2} \mathbin{\wedge}  \temporal{F} (\ap{AP1} \mathbin{\wedge}  \temporal{G}~\ap{AP1})$ we see that ``there'' must refer to \ap{AP1}.
As this type of ambiguity cannot be resolved without this hidden information, we perform minimal rewrites to the NL specification to provide a direct reference; e.g., replacing ``there'' with \ap{AP1} in this case.
In total, we edit $7$ (8.3\%) NL specifications of the \ConfoNLStrict dataset.
We analyze both the filtered and strict versions in the evaluation in \S~\ref{sec:expII}.
All dataset variants are available; see \S~\ref{sec:conclusion}.

\begin{tcolorbox}
\textbf{RQ1 answer.}
We address 
existing datasets' weaknesses 
through targeted filtering,
removing up to $48.43$\% of entries due to test-train contamination, trivial formulae,
temporal ambiguity, and \LTLf incompatibility and editing an additional $4.27$\% of the remaining entries for anaphoric ambiguity,
improving
the construct and external validity of 
\S~\ref{sec:expII} results.

\end{tcolorbox}

\subsection{RQ2: Benchmark Simplicity}
\begin{figure}[t]
    \centering

    \includegraphics[width=0.85\columnwidth,height=0.22\textheight,keepaspectratio]{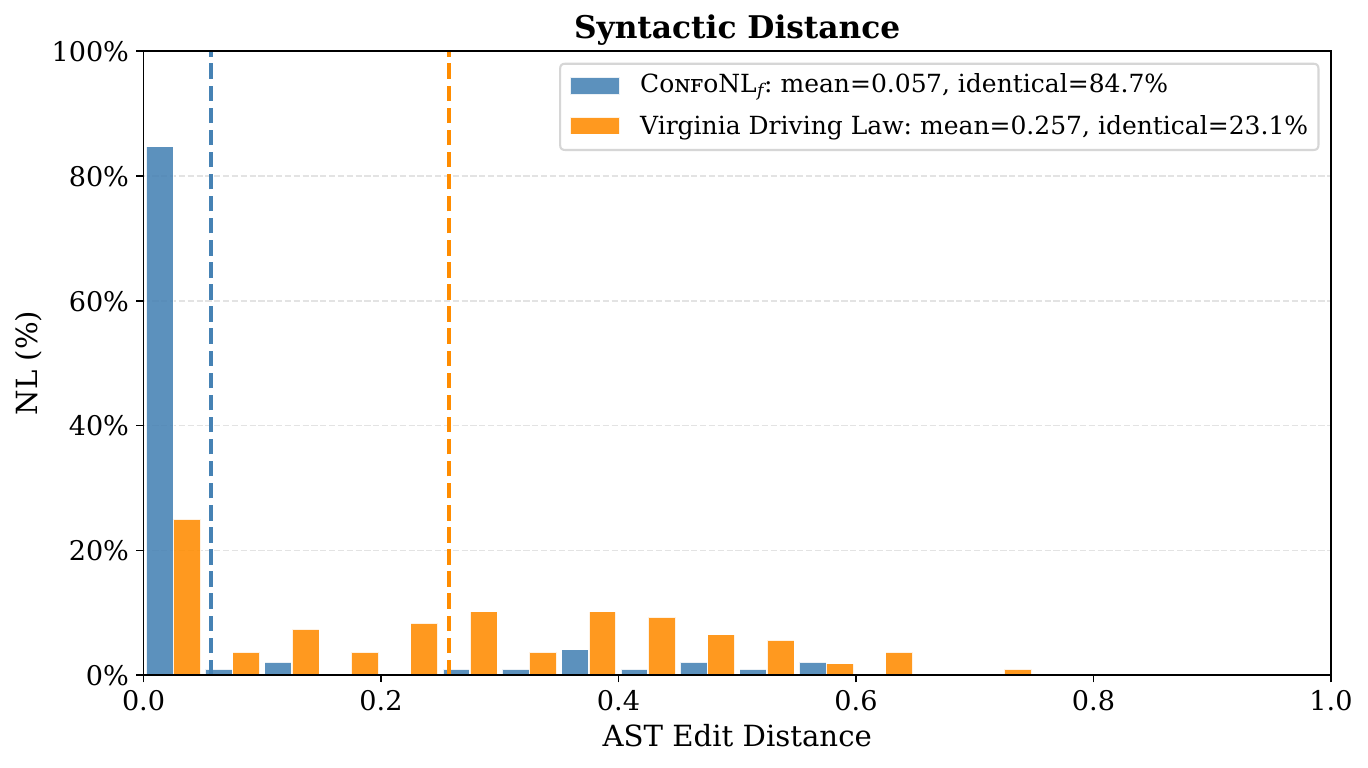}

    \vspace{0.3em}

    \includegraphics[width=0.85\columnwidth,height=0.22\textheight,keepaspectratio]{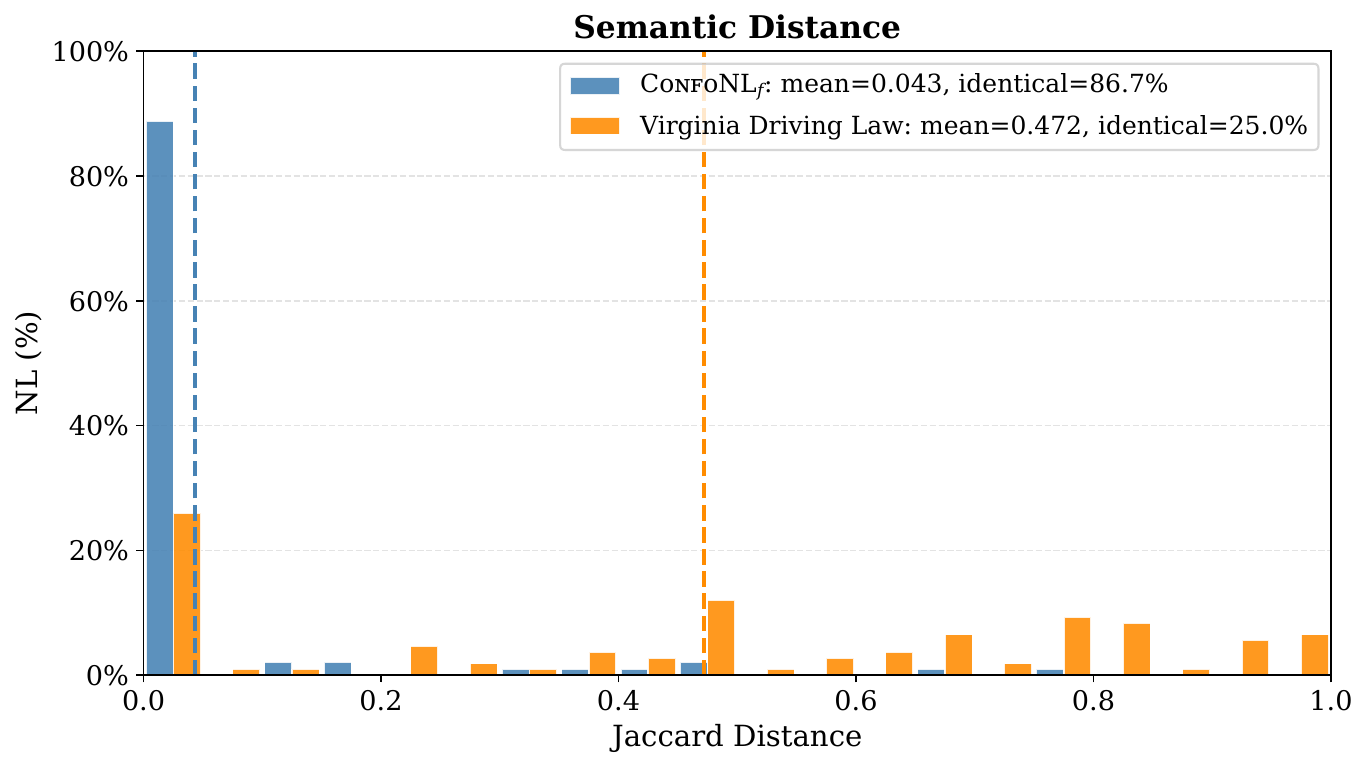}

    \caption{Distance comparison between the generated candidate formulae for \ConfoNLFiltered and Virginia driving law datasets.}
    \label{fig:diversity_comparison}
\end{figure}

To evaluate how well existing benchmarks reflect the spectrum of real-world requirements, we compare \ConfoNLFiltered\cite{sundarsingh2026conformal}, an established real-world benchmark, with the Virginia driving law (\VALaw) dataset (construction details in \S~\ref{sec:expIII_eval_design}).
While both datasets originate from real-world domains, they capture natural language specifications with different characteristics. 
First, while \ConfoNLFiltered and \VALaw have a similar number of entries at $98$ and $108$, the entries have very different lengths, averaging $9$ and $273$ words, respectively.

Second, these datasets bring very different scales of interpretive challenges.
To compare their associated difficulty, we generate \LTLf formulae from both datasets using $4$ LLM generators (detailed in \S~\ref{sec:expII_setup}).
We then measure the syntactic and semantic distances among these generated formulae.
High diversity within the candidate formulae for a single input serves as a proxy for \emph{syntactic} and \emph{semantic}
interpretive challenges
respectively, indicating that the natural language source
led to multiple conflicting interpretations.
In Fig.~\ref{fig:diversity_comparison}, we plot the syntactic and semantic distance of the candidate formulae produced by the different LLM generators for both datasets.
Here, \textit{syntactic distance} is defined as the normalized abstract syntax tree (AST) edit distance between two \LTLf formulae, and \textit{semantic distance} is the Jaccard distance between sets of traces generated by a depth-bounded traversal of the associated deterministic finite automata (DFA)~\cite{1454320}.

On \ConfoNLFiltered, the generators produce syntactically identical candidates for 84.7\% of inputs, rising to 86.7\% under semantic comparison. 
The corresponding mean syntactic and semantic distances are low, at 0.057 and 0.043, respectively, showing that existing benchmarks tend to yield clearer semantic interpretations.
In contrast, only 23.1\% of the \VALaw inputs yield syntactically identical candidates, and only 25.0\% result in semantically identical candidates.
Moreover, the mean syntactic distance increases to 0.257 
($\approx4.5\times$ \ConfoNLFiltered),
and
0.472 
($\approx11\times$ \ConfoNLFiltered) for semantic distance.

The broader \VALaw \LTLf distributions, particularly in semantic distances, indicate that these differences are not merely alternative syntactic renderings of the same behavior but often genuinely different behavioral interpretations.
This suggests that existing real-world benchmarks like \ConfoNLFiltered may lead us to underestimate the interpretive challenges present in requirements. 
The \VALaw dataset demonstrates that field specifications can induce severe variations in linguistic interpretation due to the added complexity and nuance, leading to high formula indeterminacy that current NL-to-\LTL benchmarks do not fully capture.


\begin{tcolorbox}
\textbf{RQ2 answer.} Existing benchmarks substantially underrepresent the difficulties in interpreting real-world specifications. Compared with the \VALaw dataset, \ConfoNLFiltered has on average $30\times$ fewer words per entry, and shows much lower variation among LLM-generated candidate formulae, with mean syntactic and semantic distances that are approximately $4.5\times$ and $11\times$ smaller, respectively.
\end{tcolorbox}

\section{The \NeuroSTAR Approach}

Given an
NL
requirement and an atomic proposition (AP) library, the goal of \NeuroSTAR is to formalize the NL into an \LTLf formula that is syntactically valid, AP-consistent, and semantically faithful to the NL.

The uniqueness of \NeuroSTAR lies in its neuro-symbolic integration of LLM-based generation with formal-methods reasoning.
Instead of relying on a single generated formula or only comparing formulae at the textual level, \NeuroSTAR generates multiple candidate \LTLf formulae, translates them into automata, and uses automata-theoretic operations to make their behavioral differences explicit.
These concrete trace-level differences are then used to guide refinement and allow the framework to move beyond one-shot NL-to-\LTLf generation toward a training-free iterative refinement process grounded in formal symbolic reasoning.

\subsection{Key Insights}
\label{sec:key_insight}
Recent advances in LLMs 
have made it possible to treat
NL-to-\LTLf translation as a multi-generator problem rather than relying on a single translator. Different \LTLf generators may capture different aspects of the same NL description: one candidate may be too strict, another 
too weak, and a third may better express the intended temporal ordering, thereby introducing useful diversity in the candidate formula set. Moreover, diversity can also be increased by sampling from the same LLM at different temperatures, where lower temperatures tend to produce more conservative and stable formulae and higher temperatures encourage more varied candidates that may capture alternative interpretations or edge cases.

Given the diversity of \LTLf{} formulae produced by different generators, \NeuroSTAR{} leverages this variation during the generation phase as a source of complementary semantic insight. By translating candidate formulae into DFAs and computing difference automata, the framework identifies concrete traces that expose semantic differences among the formulae. These traces are then judged against the NL description, and the resulting feedback is used to help generators refine their candidates, 
combining the strengths of different formulae and ultimately producing a final \LTLf{} formula that captures multiple perspectives and edge cases.

\begin{figure}[t]
\center
\resizebox{0.9\columnwidth}{!}{\centering
\begin{tikzpicture}[
  node distance=1.6cm,
  stage/.style={
    draw, rounded corners, thick,
    minimum width=9cm,
    inner sep=8pt
  },
  box/.style={
    draw, rounded corners,
    minimum height=6mm,
    inner sep=3pt,
    font=\scriptsize
  },
  arrow/.style={->, thick}
]

\newcommand{\stacknode}[3]{%
  \node[box, #2, opacity=0] (#1-base) {#3};

  \node[box, fill=white, text opacity=0]
    at ([xshift=-4pt,yshift=4pt]#1-base) {#3};
  \node[box, fill=white, text opacity=0]
    at ([xshift=-2pt,yshift=2pt]#1-base) {#3};

  \node[box, #2, fill=white] (#1) {#3};
}

\node[box] (nl) {NL ($x$)};
\node[box, right=10pt of nl] (ap) {AP Library $(\mathcal{L}_{AP})$};

\coordinate (join) at ($(nl.south) + (0,-2mm)$);
\draw[thick] (nl.south) -- (join);
\draw[thick] (ap.south) |- (join);

\node[below=of ap, yshift=10mm] (s1) {\textbf{Candidate $\tLTLf$ Generation and Validation}};

\stacknode{ltlf}{below=of s1, xshift=15mm, yshift=12mm}{$\tLTLf(\phi_i)$}

\node[box, below=of ltlf, yshift=10mm, fill=white] (syn)
{$\mathrm{SyntaxCheck}(\cdot) \land (\mathrm{AP}(\cdot)\subseteq \mathcal{L}_{AP})$};

\stacknode{gen}{left=of syn, xshift=5mm, yshift=5mm}{Generator $(G_i)$}

\draw[arrow] (gen.east) -| ++(5mm,0) |- (ltlf.west);
\draw[arrow] (ltlf.south) -- (syn.north);
\draw[arrow] (syn.west) -| node[pos=0.3, below] {invalid} (gen.south);
\draw[arrow] ($(gen.west)+(-8.5mm,0)$) -- (gen.west);

\begin{scope}[on background layer]
\node[
  draw,
  rounded corners,
  fill=red!10,
  inner sep=6pt,
  fit=(s1) (ltlf) (syn) (gen)
] (stage1box) {};
\end{scope}

\stacknode{valid}{below=of stage1box.south, yshift=12mm}{Valid $\tLTLf(\phi_i)$}
\draw[arrow] (syn.south) -- ++(0,-3mm) -| node[right,xshift=14mm,yshift=-1mm]{valid} (valid.north);

\node[below=of stage1box.south] (s2) {%
  \makebox[0pt][l]{\hspace{-33mm}\textbf{Automata-Based Differencing}}%
};

\node[box, below=of s2, xshift=15mm, yshift=12mm, fill=white] (ltl2dfa) {$\tLTLf$ to DFA};

\stacknode{dfas}{left=of ltl2dfa, xshift=5mm, yshift=0mm}{$D_i$}

\node[box, left=of dfas, xshift=5mm, yshift=0mm, fill=white] (diffs) {$U_i, V_i$};

\draw[arrow] (valid.east) -| (ltl2dfa.north);
\draw[arrow] (ltl2dfa.west) -- (dfas.east);
\draw[arrow] (dfas.west) -- (diffs.east);

\coordinate (stage2leftpad) at ($(diffs.west)+(-4mm,0)$);

\begin{scope}[on background layer]
\node[
  draw,
  rounded corners,
  fill=yellow!80!orange!30,
  inner sep=6pt,
  fit=(s2) (ltl2dfa) (dfas) (diffs) (stage2leftpad)
] (stage2box) {};
\end{scope}

\stacknode{kadiffs}{below=of stage2box.south, yshift=10mm}{$\mathcal{K}_i \subseteq L(U_i), \mathcal{A}_i \subseteq L(V_i)$};

\draw[arrow] (diffs.south) -- ++(0,-4mm) -| (kadiffs.north);

\node[below=of stage2box.south] (s3) {%
  \makebox[0pt][l]{\hspace{-22mm}\raisebox{3mm}{\textbf{Judging and Feedback}}}%
};

\node[box, below=of s3, xshift=-30mm, yshift=15mm, fill=white] (judge) {Judge};

\stacknode{refinements}{right=of judge, xshift=-5mm}{$R_i$, $I_i$};

\node[box, right=of refinements, xshift=-5mm, fill=white] (conv) {Convergence Checker};

\draw[arrow] (kadiffs.west) ++(0,0) -| (judge.north);
\draw[arrow] (kadiffs.east) -| ($(conv.north) + (3mm,0)$);
\draw[arrow] (join.west) -- ++(-25mm,0) |- (judge.west);

\draw[arrow] (judge.east) -- (refinements.west);
\draw[arrow] (refinements.east) -- (conv.west);

\begin{scope}[on background layer]
\node[
  draw,
  rounded corners,
  fill=blue!10,
  inner sep=6pt,
  fit=(s3) (refinements) (conv) (judge)
] (stage3box) {};
\end{scope}

\node[box, right=of conv, xshift=0mm, yshift=10mm, rotate=90] (fb) {Refinement Feedback};
\node[box, right=of conv, xshift=-5mm, align=center] (final) {Final\\$\tLTLf$};

\draw[arrow] ($(conv.north east) + (-0.5mm,-0.5mm)$) -- node[below,xshift=1.5mm]{no} (fb);
\draw[arrow] (conv) -- node[below]{yes} (final);
\draw[arrow] (fb.east) |- (stage1box.east);

\draw[arrow] (valid.east) -- ++(20mm,0)
  |- ($(conv.north)+(8mm,13.5mm)$)
  -- ($(conv.north)+(8mm,0)$);

\end{tikzpicture}
}
    \caption{\NeuroSTAR framework overview.}
    \label{fig:neuroStar}
\end{figure}

\subsection{Approach Overview}
Fig.~\ref{fig:neuroStar} summarizes the proposed \NeuroSTAR framework consisting of three modules. First, in the \textit{Candidate \LTLf Generation and Validation} module, given an 
NL
description and an
AP
library, the framework uses multiple \LTLf generators to produce candidate \LTLf formulae and checks them for syntax validity and AP consistency. 
Valid \LTLf formulae are passed to the \textit{Automata-Based Differencing} module, which compiles these into DFAs, and, using DFA operations, generates two sets of differential traces for each \LTLf candidate.
One set represents the behaviors that the \LTLf uniquely captures, and another represents the behaviors that the \LTLf misses with respect to other candidates. 
Then the \textit{Judging and Feedback} module decides which behaviors should be added and which behaviors should be removed from each \LTLf.
These decisions are converted into structured refinement feedback and are passed to the generators in the next iteration.
The loop terminates when the framework reaches convergence, as described in \S~\ref{sec:llm_judging_feedback}, or when the maximum number of iterations is reached.

\subsection{Candidate \LTLf Generation and Validation}
The input to this first module is a natural-language description, an AP library, and optional refinement feedback from the previous iteration.
The output is a set of syntactically valid and AP-consistent candidate \LTLf formulae.

First, $N$ \LTLf \textit{Generators} are used to generate candidate \LTLf formulae for the NL description.
Let $G_i$ denote generator $i$. At each iteration, generator $G_i$ receives the natural-language description $x$, the AP library $\mathcal{L}_{\mathrm{AP}}$, and refinement feedback $F_i$. 
Here, $F_i$ is structured textual refinement feedback that summarizes how the previous candidate \LTLf should be modified.
$F_i$ is empty in the first iteration, and its construction is described in \S~\ref{sec:llm_judging_feedback}.
The generator uses this refinement feedback as additional context to produce the candidate \LTLf formula, $
    \phi_i =
    G_i\!\left(x, \mathcal{L}_{\mathrm{AP}}, F_i\right)
$

Here, $\mathcal{L}_{\mathrm{AP}}$ denotes the AP library for the target domain, where each atomic proposition $p$ is paired with a natural-language description $d(p)$. 
This library is the fixed formal vocabulary over which the \LTLf formulae are expressed.
While NL specifications may describe domain events, states, or conditions in different ways, an AP library provides the grounding between the informal NL description and the formal propositions available to the generators.
Depending on the dataset, this library may be explicitly provided, implicitly embedded in the input specification, constructed by the dataset designer or a domain expert, or generated using an LLM.
In this paper, when $\mathcal{L}_{\mathrm{AP}}$ is not available, we use an LLM to generate it, with construction details provided in \S~\ref{sec:expIII_eval_design}.

Each candidate \LTLf formula must pass through two validity checks before it is used for DFA comparison.
First, the formula must be syntactically valid, meaning that it must conform to the standard \LTLf syntax.
Second, every AP referenced in the formula must belong to the AP library $\mathcal{L}_{AP}$.

We denote the set of APs that occur in formula $\phi_i$ by $\mathbf{AP}(\phi_i)$. The validity condition is then:
\begin{equation*}
\mathbf{Valid}(\phi_i) =
\mathbf{SyntaxCheck}(\phi_i)
\wedge
\left(
\mathbf{AP}(\phi_i) \subseteq \mathcal{L}_{AP}
\right)
\end{equation*}

If a formula fails the validity condition, the corresponding generator receives a message that details the error and is asked to repair the formula up to a fixed number of retries. 
This validity check removes malformed formulae and AP hallucinations before the more expensive semantic comparison.

\subsection{Automata-Based Differencing}
\label{sec:trace_gen}
The validated candidate \LTLf formulae from the previous module are passed to this module as input. The module converts each of these \LTLf formulae into a DFA, constructs difference automata that capture behavioral disagreements among candidates, and then generates traces from those automata as concrete signals of the behaviors. 

First, the \textit{\LTLf to DFA} converter converts each candidate formula $\phi_i$ into an equivalent DFA $D_i$. Then, to expose behavioral disagreements among the \LTLf formulae, we define two difference automata $U_i$ and $V_i$ as follows:

\begin{equation*}
\begin{aligned}
U_i &=
D_i \setminus
\left(
    \bigcup_{j \neq i} D_j
\right)
\qquad
V_i &=
\left(
    \bigcup_{j \neq i} D_j
\right)
\setminus
D_i
\end{aligned}
\end{equation*}

Thus, the language of $U_i$, denoted by $L(U_i)$, contains the traces accepted only by $\phi_i$ and rejected by all other candidates, and
the language of $V_i$ contains traces rejected by $\phi_i$ but accepted by one or more alternative candidates. 

From the difference automata, the framework generates two sets of witness traces where each witness is a finite accepting path in the corresponding difference automaton, rendered as a concrete trace.
\begin{equation*}
    \mathcal{K}_i \subseteq L(U_i)
    \qquad
    \mathcal{A}_i \subseteq L(V_i)
\end{equation*}

Here, the set $\mathcal{K}_i$ contains traces uniquely accepted by $\phi_i$, while $\mathcal{A}_i$ contains traces accepted by at least one other candidate but rejected by $\phi_i$.

A key design parameter of this module is the trace generation strategy used to generate finite witnesses from the difference automata. 
The trace budget, i.e., the number of traces to generate from each difference automaton $(\left|\mathcal{K}_i\right|, \left|\mathcal{A}_i\right|)$, together with the characteristics of the target domain, determines which generation strategy makes the generated traces more effective in capturing the behavior of the difference automata. 
For example, the trace generator may prioritize edge coverage, state coverage, path coverage, truth-value diversity, or diverse combinations of atomic propositions.
The appropriate choice is domain- and dataset-dependent. 
Overall, the goal is to obtain a finite set of traces that collectively captures as much of the relevant behavior of the difference automata as possible within the given budget.
The specific trace generation strategy used in our experiments is described in \S~\ref{sec:expII_setup}.

The generated witness traces provide concrete signals of semantic disagreement among the candidate \LTLf formulae and are used in the next module to guide refinement of the candidates through feedback.

\subsection{Judging and Feedback}
\label{sec:llm_judging_feedback}
The input to this module is the NL specification $x$, the AP library $\mathcal{L}_{AP}$, the candidate \LTLf formulae from the first module, and the witness traces $\mathcal{K}_i$ and $\mathcal{A}_i$ generated in the second module.
Optionally, the module may also receive the same NL--\LTLf{} example pairs provided to the generators as additional dataset-specific context for the judges. The output is either a final \LTLf formula $\phi^*$ or structured refinement feedback $F_i$ for the next iteration. 

This module uses LLM judges to decide whether the generated trace behaviors are intended by the NL specification, converts those decisions into refinement feedback, and checks whether a candidate has converged.
The module is designed around the difference in difficulty between synthesizing a complete formal specification and judging concrete trace-level behaviors. Directly translating natural language into a syntactically valid, AP-consistent, and semantically faithful \LTLf formula requires constructing a complete formal specification, which is difficult for LLMs. 
In contrast, judging whether a concrete finite trace is consistent with an NL specification is a more localized and interpretable task that shifts the LLM’s role from formal synthesis to semantic judgment, where its language-understanding capabilities are better suited.

Here, from the two sets of traces $\mathcal{K}_i$ and $\mathcal{A}_i$, we aim to extract traces that are consistent with the NL specification. 
We use two panels of LLM judges 
for this task.

First, for each trace in $\mathcal{K}_i$, a panel of LLM judges, cumulatively defined as $Judge_{keep}$, is asked to describe the trace and judge whether the trace is an intended behavior of the NL specification $x$ and the generator $i$ should keep the trace. Each judge in $Judge_{keep}$ outputs a binary decision, a trace description, and a reasoning of the decision. The final decision for each of the traces is taken using majority voting on the binary decisions, and all corresponding trace descriptions and reasoning of that binary decision are concatenated into one.

Similarly, for each trace in $\mathcal{A}_i$, a panel of judges, $Judge_{add}$, describes the trace and determines whether the behavior represented by the trace should be added to the candidate \LTLf formula, based on its alignment with the NL specification.

Here, a \textit{YES} decision from $Judge_{keep}$ means the unique trace appears consistent with the NL and should be kept by the corresponding generator. 
Conversely,
a \textit{NO} decision means this trace is not a part of the intended behavior, and the generator should remove it from the formula in the next iteration. 

Similarly, a \textit{NO} decision from $Judge_{add}$ denotes that the missing trace should not be added because it is inconsistent with the NL, a \textit{YES} decision here signifies that the candidate is missing this intended behavior and should be relaxed or repaired to incorporate it. We cache and reuse judge decisions for optimization and consistency.

Concretely, the judge decisions produce a set of traces $\mathcal{R}_i$ that should be removed from $\phi_i$ and a set of traces $\mathcal{I}_i$ that should be inserted into $\phi_i$ in the next iteration.
Based on these traces, we define two stability scores for $\phi_i$:

\[
J_{\mathcal{K}}(\phi_i)=
(\lvert\mathcal{K}_i\rvert-\lvert\mathcal{R}_i\rvert)/\lvert\mathcal{K}_i\rvert
\qquad
J_{\mathcal{A}}(\phi_i)=
(\lvert\mathcal{A}_i\rvert-\lvert\mathcal{I}_i\rvert)/\lvert\mathcal{A}_i\rvert
\]

Here, high $J_{\mathcal{K}}(\phi_i)$, $J_{\mathcal{A}}(\phi_i)$ scores signify that $\phi_i$ is locally stable relative to others, i.e., it does not need many local modifications. Using these stability scores, the \textit{Convergence Checker} determines whether the output has converged. First, from the full set of candidate \LTLf formulae, the \textit{Convergence Checker} selects the subset $\Phi_{\mathrm{stable}}$ that satisfies the following:

\begin{equation*}
    \Phi_{\mathrm{stable}}
    =
    \{ \phi_i \mid
    J_{\mathcal{K}}(\phi_i) \geq m
    \wedge
    J_{\mathcal{A}}(\phi_i) \geq n
    \}
\end{equation*}

If fewer than three candidates satisfy these conditions, i.e., $|\Phi_{\mathrm{stable}}| < 3$, the checker does not attempt convergence and proceeds to the next iteration. Otherwise, it computes all pairwise Jaccard distances~\cite{1454320} among the formulae in $\Phi_{\mathrm{stable}}$. Let $d_J(\phi_i,\phi_j)$ denote the bounded trace Jaccard distance between two formulae $\phi_i$ and $\phi_j$. 
Using this distance, the checker then selects the medoid~\cite{baharav2019ultrafastmedoididentification},
defined as the formula with the smallest distance to the other stable candidates:

\begin{equation*}
\phi^*
=
\arg\min_{\phi_i \in \Phi_{\mathrm{stable}}}
\sum_{\phi_j \in \Phi_{\mathrm{stable}}}
d_J(\phi_i,\phi_j)
\end{equation*}

The candidates are considered converged if the medoid 
is within a distance threshold $\theta$ of 
another stable candidate:

\begin{equation*}
\min_{\substack{\phi_j \in \Phi_{\mathrm{stable}} \ \phi_j \neq \phi^*}}
d_J(\phi^*,\phi_j)
\leq \theta
\end{equation*}

Intuitively, convergence here means that the selected formula is stable under judge feedback, i.e., it requires few trace-level corrections, and it is also semantically close to another separately generated stable candidate. Thus, the final output needs both semantic stability and generator agreement. If the convergence condition is satisfied, the medoid formula $\phi^*$ is returned as the final output formula.

Alternatively, a set of outputs is considered to be converged when all of them are semantically equivalent. In that case, \(L(U_i) = L(V_i) = \emptyset\) for every candidate \(\phi_i\), indicating that no distinguishing trace exists between any candidate and the others, and both the judge checks and the Jaccard distance checks are vacuously satisfied. 
In some cases, all generators can produce the same wrong formula together; however, we intentionally converge in this case rather than adding any additional checks because empirical evaluation showed the likelihood of all models giving the same wrong answer was much less than their likelihood of being correct together.

Finally, if the outputs have not converged, for each candidate formula $\phi_i$, the judges' trace-level decisions $\mathcal{R}_i$, $\mathcal{I}_i$, their descriptions, and reasoning are passed to its generator as refinement feedback $F_i$.
This feedback
also includes the anonymized full set of candidate formulae
without revealing the LLM or generator identities, 
removing
model-specific bias and explicitly targeting the trace-level weaknesses identified for $\phi_i$ in the next iteration.
It includes the full candidate set, as the finite distinguishing traces alone may not uniquely determine an appropriate global \LTLf repair. 
Therefore, providing the full set lets the generator avoid overfitting to individual traces and move toward a generalized repair of the \LTLf.
Each generation round uses a new context window to ensure that the generator has no history of its previous outputs.

\section{Technique Evaluation}
\label{sec:expII}
We
evaluate whether \NeuroSTAR improves NL-to-\LTLf translation when compared to existing baseline techniques. Concretely, we answer the following research question:

\textbf{RQ3:} How does \NeuroSTAR perform against baselines on existing benchmarks?

\subsection{Evaluation Design}
\label{sec:expII_setup}
To instantiate \NeuroSTAR, we need multiple \LTLf generators, two panels of judges, and a trace generation strategy.

For each input, \NeuroSTAR uses four LLM generators: Gemini 3.5 Flash~\cite{gemini35flash} at temperatures $0.1$ and $0.9$, and Kimi K2.5~\cite{kimik25} at temperatures $0.1$ and $0.9$ with 5 few-shot examples taken from the \AdaRule train split.
By varying both the model family and sampling temperature, we let the generators produce a diverse set of candidate formulae.
We use Kimi K2.5
at temperature $0.1$ as both the $Judge_{keep}$ and $Judge_{add}$ along with the same examples provided to the generators as additional context.
However, our framework is flexible enough to support a larger set of generators and judges.

After the generation and validation phase, following \S~\ref{sec:trace_gen}, \NeuroSTAR converts each candidate formula to a DFA using the Spot Python library~\cite{duret.22.cav} and computes the difference automata $U_i$ and $V_i$, from which it derives the two witness trace sets $\mathcal{K}_i$ and $\mathcal{A}_i$ for every candidate $\phi_i$.
To generate these witness trace sets, it uses a strategy that maximizes the edge coverage of the difference DFAs. 
First, \NeuroSTAR generates all accepted traces of length $\le l_C$ to have an initial pool of traces $\mathcal{C}$.
Then, given a trace budget $b$, the algorithm 
selects
at most $b$ traces from $\mathcal{C}$ whose combined traversed-edge set is maximal.
\NeuroSTAR encodes this maximum edge-coverage objective as a mixed-integer linear program and uses the \textit{HiGHS MILP} solver~\cite{huangfu2018parallelizing} 
in SciPy~\cite{2020SciPy-NMeth} to generate an optimal solution.
The selected subset is therefore globally optimal over the enumerated candidate trace set, rather than a greedy approximation.
We set $l_C = 5$ and $b = 5$, values chosen empirically from pilot runs that measured the size of the difference DFAs and their accepted traces.
These values provide high behavioral coverage of the difference DFAs while keeping the trace count low to manage the size of the judge prompts. 
For RQ3, we evaluate over the filtered and the strict versions of the four datasets
described in \S~\ref{sec:ambiguous_nl}.


\subsection{Baselines}
We compare \NeuroSTAR with three SoTA baselines.

\paragraph{\textbf{\AdaRule}}
\AdaRule~\cite{hu2026adarule} is a guideline-based NL-to-\LTLf baseline that retrieves training-derived translation guidelines via cosine similarity between test and train NLs and injects them into the prompt during inference. As it previously showed good performance with Gemini, we used 
Gemini 3.5 Flash
with the default 16 guidelines in its inference.

\paragraph{\textbf{LfF}}
LfF~\cite{10684640} is a learning-from-failures baseline that retrieves stored NL-to-\LTL examples via cosine similarity and includes them in the prompt during inference. 
Again, we used Gemini 3.5 Flash as suggested in their method.

\paragraph{\textbf{$n$-shot prompting}}
Finally, given the NL specification and AP library, we generate the \LTLf formula via single-LLM few-shot prompting, using the same LLMs, temperatures, and number of few-shot examples ($n=5$) as the \NeuroSTAR generators, and report the average score.
This baseline isolates the benefit of \NeuroSTAR{}'s multi-candidate automata-feedback loop over ordinary few-shot generation.

We observe that no baseline examines the semantics of the generated \LTLf, instead relying on NL textual patterns.
This is the principal difference between them and \NeuroSTAR, which compares \LTLf formulae by their actual behavior.

\subsection{RQ3 Results}
\begin{table}[t]
  \centering
  \caption{Results (\%) on the filtered and strict datasets.}
  \label{tab:results_RQ3}
  \resizebox{\columnwidth}{!}{%
    \begin{tabular}{lcccc}
      \toprule
      \multicolumn{5}{c}{\textbf{Filtered version}} \\
      \cmidrule(lr){1-5}
      Method & \ConfoNLFiltered & \SynthNLFiltered & \SpecNLFiltered & \LangNLFiltered \\
      \midrule
      \AdaRule~\cite{hu2026adarule}                & 77.6          & \textbf{78.9} & \textbf{92.9} & \textbf{94.1} \\
      LfF~\cite{10684640}                      & 68.4          & 43.2          & 85.7          & 76.5 \\
      $n$-shot prompting           & 86.2          & 76.5          & 87.5          & 89.7 \\
      \NeuroSTAR & \textbf{87.8} & 75.8          & \textbf{92.9} & \textbf{94.1} \\
      \bottomrule
    \end{tabular}%
  }

  \medskip

  \begin{tabular}{lcc}
    \toprule
    \multicolumn{3}{c}{\textbf{Strict version}} \\
    \cmidrule(lr){1-3}
    Method & \ConfoNLStrict & \SynthNLStrict \\
    \midrule
    \AdaRule~\cite{hu2026adarule}                    & 88.1          & 81.6 \\
    LfF~\cite{10684640}                    & 75.0          & 75.5 \\
    $n$-shot prompting           & 93.4          & 97.4 \\
    \NeuroSTAR & \textbf{96.4} & \textbf{100.0} \\
    \bottomrule
  \end{tabular}
\end{table}
Table~\ref{tab:results_RQ3} presents the accuracy of \NeuroSTAR and the baselines, measured as the percentage of generated \LTLf formulae that are semantically equivalent to the corresponding ground-truth formulae calculated using Spot~\cite{duret.22.cav}.
\NeuroSTAR achieves the best or joint-best accuracy on three of four datasets in the filtered version (\ConfoNLFiltered 87.8\%, \SpecNLFiltered 92.9\%, \LangNLFiltered 94.1\%) and outperforms all baselines on both datasets in the strict version (\ConfoNLStrict 96.4\%, \SynthNLStrict 100\%).
On \ConfoNLFiltered, \NeuroSTAR surpasses the second-best method, $n$-shot prompting, by 1.6 percentage points (pp), and \AdaRule by 10.2 pp.
A similar trend appears in \ConfoNLStrict, where \NeuroSTAR exceeds $n$-shot prompting by 3.0 pp and \AdaRule by 8.3 pp. 

\AdaRule, by contrast, achieves the highest accuracy in \SynthNLFiltered, reaching 78.9\% and surpassing \NeuroSTAR by 3.1 pp. \SynthNL~\cite{crick2023nl2ltlsynthetic} is a synthetic dataset constructed from eight predefined specification patterns. Since all of its NL-\LTLf pairs are drawn from these eight patterns, a training-based method such as \AdaRule can easily memorize these patterns through its guideline generation and score well here. Such memorization, however, transfers poorly to realistic datasets like \ConfoNL~\cite{sundarsingh2026conformal}, which consists of real-world robot task instructions. Moreover, once the ambiguous patterns are removed in \SynthNLStrict, \NeuroSTAR achieves the highest performance of 100\%, showing that even learned pattern-based methods can not match \NeuroSTAR when the examples are unambiguous.


We also find that \AdaRule and LfF sometimes fail to extract the correct AP names from the NL descriptions. In contrast, \NeuroSTAR and $n$-shot prompting explicitly take the AP library as input, which helps them produce formulae with valid AP names. On top of that, \NeuroSTAR includes an AP-validity check, which is essential because using incorrect APs immediately invalidates the generated \LTLf formula.

\begin{tcolorbox}
\textbf{RQ3 answer.}
\NeuroSTAR outperforms the baselines on realistic and unambiguous benchmarks, demonstrating the effectiveness of its automata-guided refinement.
\end{tcolorbox}

\section{Proof-of-Concept Study}
\label{sec:va-law}
Here, we analyze the performance of \NeuroSTAR on realistic driving-law text where reference \LTLf formulae are unavailable. 
We answer
the following research question:

\textbf{RQ4:} How well does \NeuroSTAR perform on real-world \VALaw dataset?

\subsection{Evaluation Design}
\label{sec:expIII_eval_design}
We selected the Virginia driving code~\cite{virginia_code_title46_2_2025} as a real-world NL corpus to study for several reasons.
First, it was written by domain experts, including transportation experts, lawyers, and government officials, to capture constraints on vehicles and drivers.
Second, it reflects the breadth, diversity, and complexity of a rich problem domain.
From Fig.~\ref{fig:diversity_comparison}, one can observe that both the syntactic
and semantic diversity is much less concentrated than prior benchmarks.
Finally, prior work had already done an initial triage 
of the 207 numbered sections to identify 114 that are applicable to autonomous driving systems~\cite{woodlief2025scene}, and had gone through the manual effort to create \LTLf
specifications for 6 of them~\cite{WOODLIEF2025103252}. 
In this study, we focus on the remaining 108 sections that have no ground truth.

This study differs from the other evaluations of \NeuroSTAR in important ways. 
First, no benchmark exists for the Virginia driving laws, so we web-scrape the raw NL text of each law to construct the dataset ourselves. 
Second, no AP library is available for this dataset.
To address this, we use the 6 manually created \LTLf formulae~\cite{WOODLIEF2025103252} as few-shot examples and prompt Claude Sonnet 4.6~\cite{claudesonnet46} with the full set of driving laws to generate a library
of 352 APs that reflect the atomic behavior mentioned across the 114 sections of the driving code. 
Third, sections contain cross-references, where a provision depends on definitions or exceptions described elsewhere. 
Cross-references use an idiomatic syntax in the driving code of the form ``\S~46.2-n.m''
where ``n'' is the section number and ``m'' the subsection number.  
This allows us to identify cross-references and provide the referenced section as additional
context for \NeuroSTAR.
Fourth, given the 
increased length of \VALaw specifications,
we use $l_C=10$ and $b=10$; 
we also use the existing 6 \LTLf as few-shot examples.
Finally, unlike the previous datasets, no reference \LTLf formulae exist. 
This makes comparison with training-dependent approaches, such as \AdaRule,
impossible on this dataset.
We discuss this issue in the next section.

\subsection{Evaluation Metric}
\label{sec:exp_studyIII_metric}
The lack of ground truth \LTLf led us to develop a metric that (a) provided a
signal about the correctness of the generated specification, and (b) was not 
cost-prohibitive.
We found that comparing the 
boolean
and temporal structure of an 
\LTLf from NL text that was on average $273$ words long was challenging and took
significant time.
This led us to shift to considering traces, which allow
one to navigate a ``path'' through the NL phrasing to determine
a correspondence.

More specifically, we leverage the symbolic DFA representations of generated \LTLf to
generate violating traces---sequences of 
AP valuations that lead to a violation.
This forms a necessary condition for the \LTLf to be correct, i.e., if a violating
trace of the \LTLf is not a violation of the NL, then the \LTLf
is wrong. 
Two authors were randomly assigned to each driving code section
and tasked with assessing whether the violating trace computed for the generated
\LTLf also violated the section's intent as expressed in NL.
The rating protocol stated that a rater should be conservative and indicate
that it was a violation only if they were sure that was the case.
When the raters disagreed, a third was assigned.
The majority decision defines the following metric.
\begin{figure}[t]
    \centering
    \includegraphics[width=0.9\columnwidth]{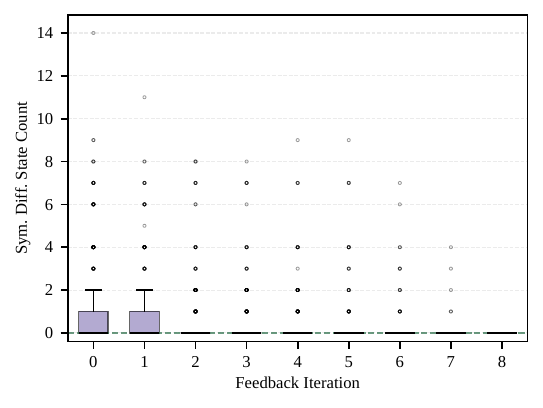}
    \caption{Cumulative semantic convergence of generated \LTLf formulae to the final converged formulae across iterations.}
    \label{fig:convergence_fig}
\end{figure}

Given
an NL specification from
the driving code, $x$, and an
\LTLf, $\phi$, generated from $x$, let $w$ be a violating trace 
with the shortest length, or selected randomly from equally shortest traces.
If a majority of three conservative human raters judge $w$ to be a violation
of $x$ by interpreting the specification and AP semantics,
then we say the \textbf{violating witness is valid} (VWV).

\begin{figure*}[t]
\centering
\footnotesize
\begin{tcolorbox}[colback=white, colbacktitle=blue!5!white, coltitle=black!55!black,
  colframe=blue!55!black, boxrule=0.5pt, arc=2pt,
  left=7pt, right=7pt, top=5pt, bottom=5pt, fonttitle=\bfseries,
  title={\S~46.2-829 --- Yielding to an approaching emergency vehicle}]
  \vspace{-2mm}
\textbf{Summary}: 
When an emergency vehicle approaches giving an audible siren signal and
displaying emergency lights, the driver of every other vehicle shall move to
the nearest edge of the roadway, clear of any intersection, and stop and remain
there---\emph{unless} otherwise directed by a law-enforcement officer---\emph{until}
the emergency vehicle has passed.

\medskip
\textbf{NeuroSTAR LTL\textsubscript{$f$} formalization}:
$
\temporal{G}\,{\color{blue}(}{\color{red}(}\,\phi_1 \land \lnot\,\ap{officer\_directing\_traffic}\,{\color{red})} \;\rightarrow\; \temporal{F}\,{\color{red}(}\,\phi_2 \land {\color{teal}(}{\color{orange}(}\,\phi_2~\mathbin{\temporal{U}} ~\phi_3\,{\color{orange})} \lor \temporal{G}\,\phi_2\,{\color{teal})}{\color{red})}{\color{blue})}
$

\medskip
where $
\begin{aligned}[t]
\phi_1 &= \ap{v1\_is\_emergency} \land \ap{v1\_emergency\_lights} \land \ap{v1\_emergency\_siren} \land \ap{v1\_approaching}\\
\phi_2 &= \ap{ego\_stopped} \land \lnot\,\ap{ego\_at\_junc} \land {\color{blue}(}\,\ap{ego\_on\_shoulder} \lor \ap{ego\_in\_rightmost\_lane}\,{\color{blue})}\\
\phi_3 &= \lnot\,\phi_1 \;\lor\; \ap{officer\_directing\_traffic}
\end{aligned}
$
  \vspace{-2mm}

\end{tcolorbox}
\caption{A \NeuroSTAR-generated \LTLf formula for \textit{\S~46.2-829} of \VALaw, illustrating several recurring formalization strengths. The original law description is replaced by its summary for readability.}
\label{fig:law-829}
\end{figure*}

While informative, it is important to note that VWV does not provide a signal
that the full set of accepted or rejected traces of the \LTLf is consistent with the
NL description.
That would have been cost-prohibitive.  
We view this as an initial means of assessing
\NeuroSTAR's utility in generating \LTLf formulae from real-world datasets. 
Moreover, the approach can be scaled with additional traces, both accepting
and rejecting, to provide further confidence in the assessment for future
evaluation of reference-free techniques.

\subsection{RQ4 Results}
Of the 108 laws, 107 converge, and Fig.~\ref{fig:convergence_fig} illustrates this
convergence by measuring the difference
between the final \LTLf and the \LTLf generated on the $i$th iteration (indicated on the x-axis).
Difference is measured in the number of
states of the difference automata.
By iteration 8, all 107 reach semantic equivalence with the final formula, demonstrating that iterative feedback drives effective refinement.

Among the converged 107, 1 is trivially true. Of the remaining 106, we find that 89 of the violating witnesses were valid (83.9\%), meaning that the majority of the reviewers think that the violating witness generated from \NeuroSTAR's derived \ltlf 
constitutes
a violation of the NL specification.

Fig.~\ref{fig:law-829} shows a representative example, section \textit{\S~46.2-829}, 
that
exhibits several strengths of \NeuroSTAR.
In the corresponding \LTLf, the eventually operator $\temporal{F}$ encodes ``as quickly as traffic and highway conditions permit'' and the weak-until expansion $(\phi_2 \mathbin{\temporal{U}} \phi_3)\lor\temporal{G}\,\phi_2$ enforces ``remain until the vehicle has passed'' without assuming the release $\phi_3$ ever occurs.
The officer AP appears both as a negated antecedent guard and the release disjunct, $\phi_3$, so the obligation defers to an officer whether one is directing before or during the stop. 
In the following, as part of a qualitative analysis to extract recurring patterns where \NeuroSTAR excels and identify the reasons behind its limitations, we discuss $7$ key strengths of \NeuroSTAR, $5$ of which can be observed in Fig.~\ref{fig:law-829}.

(1)~``Unless'' or ``except'' clauses, which occur at least once in 22.6\% and 30.2\% of the sections, respectively, are correctly modeled as negated antecedent literals.  In Fig.~\ref{fig:law-829},  ``unless otherwise directed by a law-enforcement officer'' yields $\neg\,\ap{{officer\_directing\_traffic}}$.
Similar constructions are observed across the corpus, e.g., $\neg\,\ap{{traffic\_control\_device\_present}}$ for ``unless otherwise directed by traffic control devices'' in \textit{\S~46.2-807}.

(2)~Laws mentioning the obligation to reach and maintain a state until a condition occurs in 11.3\% of the sections.  These are expressed using the $\mathbin{\temporal{U}}$ operator, as we see in Fig.~\ref{fig:law-829}; the driver must remain stopped at the roadway edge ``until the emergency vehicle has passed.''
A similar pattern appears in \textit{\S~46.2-859}, where drivers must stop while a school bus is loading or unloading passengers.

(3)~\NeuroSTAR discriminates among vehicle classes at the granularity the law mentions. In Fig.~\ref{fig:law-829}, the restriction is placed specifically on the counterpart being an emergency vehicle through $\ap{v1\_is\_emergency}$, leaving the other vehicle class unconstrained. Similarly, law \textit{\S~46.2-839} discriminates among multiple vehicle classes: bicycle, moped, and EPAMD.

(4)~\NeuroSTAR also captures fine-grained operational conditions.  In Fig.~\ref{fig:law-829}, the required scenario is scoped to a particular roadway condition: off the intersection and at the nearest edge, encoded as
$\neg\,\ap{{ego\_at\_junc}}$
and 
$(\ap{ego\_on\_shoulder} \mathbin{\lor} \ap{ego\_in\_rightmost\_lane})$.
This was observed in other laws, e.g., \textit{\S~46.2-803.1} included commercial-vehicle lane restriction and work-zone provisions. 

(5)~\NeuroSTAR is effective in laws involving simultaneous interactions between multiple road users, e.g., multiple vehicle interaction as in Fig.~\ref{fig:law-829} and vehicle-pedestrian interaction in law \textit{\S~46.2-924}. 

(6)~We observe \NeuroSTAR isolates the operative core of a law and builds the \LTLf around it, discarding unnecessary context surrounding it.
For example, \textit{\S~46.2-873} restricts speeding in a school zone, along with several administrative clauses on speed sign placement distances, which authority furnishes the signs, fines associated with speeding, etc. 
\NeuroSTAR discards these non-operative clauses and generates the \LTLf focusing on the single enforceable rule that a driver must not exceed the speed limit while in the school zone. 

(7)~Finally, we find that a key strength of \NeuroSTAR is its ability to capture several necessary and independent requirements and edge cases under the same section and join these globally scoped subformulae using conjunctions. For example, in \textit{\S~46.2-908.1}, the \LTLf splits cleanly into the independent obligations stated
by the law: an EPAMD speed limit, a scooter speed limit, a minimum-age restriction, and an interstate-highway prohibition,  each expressed as a separate globally scoped conjunct.

By comparison, existing benchmarks studied in \S~\ref{sec:ambiguous_nl} are much simpler, both linguistically and structurally, and use no more than $5$ APs per specification.

Across this proof-of-concept study, the dominant limitation we observed in formalizing the \VALaw dataset is the AP vocabulary gap, where the propositions that the law needs are absent from the AP library. For example, law \textit{\S~46.2-868.1} describes multiple aggressive driving scenarios: speeding, following too closely, 
ignoring lane markings,
etc. However, the generated \LTLf formula compresses the individual scenarios into a single AP \ap{ego\_aggressive\_driving} and the \LTLf becomes $\temporal{G}(\neg\,\ap{ego\_aggressive\_driving)}$. Consequently, the \LTLf formula expresses aggressive driving only at an abstract level, without representing the individual scenarios. This limits its usefulness for downstream tasks, such as test generation for autonomous-driving agents, where tests for each individual scenario are generated separately~\cite{saha2026stadaspecificationbasedtestingautonomous}. Additionally, continuous thresholds (feet, miles per hour, fines) and place-specific scopes (named routes, enumerated counties) are often collapsed because the library exposes no AP for them. 
Future work is needed on generating or
refining APs to form an adequate basis for
capturing a rich corpus of requirements like \VALaw.

\begin{tcolorbox}
\textbf{RQ4 answer.}
\NeuroSTAR is able to capture necessary temporal semantics of
83.9\% of the sections of the
\VALaw that were evaluated.
Despite observed limitations relative to the AP library, \NeuroSTAR demonstrates the ability to translate the complex structure of real-world requirement statements to \LTLf.
\end{tcolorbox}

\section{Threats to Validity}

\noindent\textbf{Internal Validity.} 
A threat to internal validity comes from potential bugs in our tool implementations, including formula parsing, automata construction, and trace generation. 
Though we test all components rigorously and manually inspect all intermediate outputs, implementation errors may still remain and affect the reported results.
Our method depends on several design choices, including convergence threshold, judge-score thresholds, trace bound, and trace budget, and reported results may vary by changing these.
To mitigate this, we empirically select these parameter values using pilot runs whenever possible.
Another threat can come from dataset cleaning, where additional errors may remain.
The datasets are manually reviewed by four researchers, including three experts in temporal logic, and to support transparency and reproducibility, we release the cleaned datasets and our annotations.
Finally, the authors' bias in the manual evaluation of \VALaw natural language and the corresponding violating witness of RQ4 constitute another threat to validity, which we mitigate by assigning $VWV=0$ whenever we are unsure about the trace violation and also resolving conflicts by a third reviewer. We release these judgments for transparency; see \S~\ref{sec:conclusion}.

\noindent\textbf{Construct Validity.} The Jaccard distance metric we use in \S~\ref{sec:llm_judging_feedback} approximates semantic equivalence between two \LTLf formulae by comparing their acceptance behavior over a finite set of bounded traces (up to length $k$). 
A potential threat is that using too few traces may cause semantically different formulae to appear similar and incorrectly converge.
To mitigate this threat, we empirically analyze how the Jaccard distance changes as the trace bound $k$ increases for both the \AdaRule and \VALaw datasets and use the value of $k$ where it stabilizes and reaches the saturation point.
This ensures that the selected trace bound provides sufficient coverage for \LTLf formulae comparison and reduces the risk of spurious convergence caused by an inadequate trace set. 

\noindent\textbf{External Validity.}
Choosing a different set of models for generators and judges may introduce variation in the reported results, which is expected in all LLM-based methodologies.
We reduce this risk by using the latest available versions of SoTA models such as Gemini and Kimi, for both our method and the baselines.
Since we do not have any learning components, our method will only get better as the LLMs improve. 
Another dataset-specific threat is that existing benchmarks may not fully capture the complexity of realistic specifications, as our analysis shows that prior datasets are often shorter and simpler.
We mitigate this threat by evaluating \NeuroSTAR on the complex and realistic \VALaw dataset.

\section{Conclusion}
\label{sec:conclusion}

In this paper, we present \NeuroSTAR, a neuro-symbolic framework for formalizing natural language specifications into \LTLf formulae. Unlike prior single-generator approaches that drive \LTLf generation through textual pattern analysis of the input NL, \NeuroSTAR employs multiple generators and exploits semantic differences among the generated candidates through automata-theoretic reasoning, guiding the generators toward a converged final formula.

Our evaluation consists of three studies. 
First, we identify ambiguities and several inconsistencies in the benchmark datasets, remove $48.43$\% of the data, and also compare the benchmarks against a real-world dataset in terms of their interpretive challenges.
These inconsistencies in existing benchmarks further motivate us for a reference-free NL-to-\LTLf framework.
Second, in the evaluation of \NeuroSTAR on existing benchmarks, we demonstrate that \NeuroSTAR achieves the best or joint-best accuracy on three of four filtered datasets and outperforms all baselines on both strict datasets, reaching $96.4\%$ on \ConfoNLStrict and $100\%$ on \SynthNLStrict, demonstrating the value of automata-guided refinement and establishing a new state of the art in NL-to-\LTLf formalization.
Third, on a realistic and reference-free \VALaw dataset, we show that \NeuroSTAR 
captures necessary temporal semantics of
$89$ of $106$ laws demonstrating its applicability in a far more complex domain with many downstream applications such as safety property monitoring and test generation. Our qualitative analysis further shows that \NeuroSTAR reliably models each law's operative core along with its temporal, conditional, and agent-specific structure.

Finally, our study reveals the AP vocabulary gap as the dominant limitation in driving law formalization. Future work can build on this direction by improving AP library construction, which may further increase \NeuroSTAR's effectiveness on complex real-world specifications.


\bibliographystyle{IEEEtran}
\bibliography{references}

\end{document}